\documentclass[sigconf]{acmart}
\AtBeginDocument{%
  }

\copyrightyear{2025}
\acmYear{2025}
\setcopyright{cc}
\setcctype{by}
\acmConference[RecSys '25]{Proceedings of the Nineteenth ACM Conference on Recommender Systems}{September 22--26, 2025}{Prague, Czech Republic}
\acmBooktitle{Proceedings of the Nineteenth ACM Conference on Recommender Systems (RecSys '25), September 22--26, 2025, Prague, Czech Republic}\acmDOI{10.1145/3705328.3748077}
\acmISBN{979-8-4007-1364-4/2025/09}

\usepackage{colortbl} 
\usepackage{graphicx} 
\usepackage{booktabs} 
\usepackage{tcolorbox} 
\usepackage{multirow}
\usepackage{makecell}
\usepackage{cleveref}
\usepackage{booktabs}
\usepackage{subfigure}
\usepackage{enumitem}
\usepackage{arydshln}
\usepackage[font=small,labelfont=bf]{caption}
\newcommand{\ie}{\textit{i.e.,~}}
\newcommand{\eg}{\textit{e.g.,~}}
\crefname{section}{\S}{\S}
\crefname{subsection}{\S}{\S}
\usepackage{threeparttable} 

\begin{document}

\title{LLM-RecG: A Semantic Bias-Aware Framework for Zero-Shot Sequential Recommendation}

\author{Yunzhe Li }
\email{yunzhel2@illinois.edu}
\orcid{1234-5678-9012}
\affiliation{%
  \institution{University of Illinois, Urbana-Champaign}
  \city{Champaign}
  \state{Illinois}
  \country{USA}
}

\author{Junting Wang}
\email{junting3@illinois.edu}
\affiliation{%
  \institution{University of Illinois, Urbana-Champaign}
  \city{Champaign}
  \state{Illinois}
  \country{USA}
}

\author{Hari Sundaram}
\email{hs1@illinois.edu}
\affiliation{%
  \institution{University of Illinois, Urbana-Champaign}
  \city{Champaign}
  \state{Illinois}
  \country{USA}
}

\author{Zhining Liu}
\email{liu326@illinois.edu}
\affiliation{%
  \institution{University of Illinois, Urbana-Champaign}
  \city{Champaign}
  \state{Illinois}
  \country{USA}
}
\renewcommand{\shortauthors}{Li et al.}

\begin{abstract}
Zero-shot cross-domain sequential recommendation (ZCDSR) enables predictions in unseen domains without additional training or fine-tuning, addressing the limitations of traditional models in sparse data environments. Recent advancements in large language models (LLMs) have significantly enhanced ZCDSR by facilitating cross-domain knowledge transfer through rich, pretrained representations. Despite this progress, domain semantic bias—arising from differences in vocabulary and content focus between domains—remains a persistent challenge, leading to misaligned item embeddings and reduced generalization across domains.

To address this, we propose a novel semantic bias-aware framework that enhances LLM-based ZCDSR by improving cross-domain alignment at both the item and sequential levels. At the item level, we introduce a generalization loss that aligns the embeddings of items across domains (inter-domain compactness), while preserving the unique characteristics of each item within its own domain (intra-domain diversity). This ensures that item embeddings can be transferred effectively between domains without collapsing into overly generic or uniform representations.
At the sequential level, we develop a method to transfer user behavioral patterns by clustering source domain user sequences and applying attention-based aggregation during target domain inference. We dynamically adapt user embeddings to unseen domains, enabling effective zero-shot recommendations without requiring target-domain interactions.

Extensive experiments across multiple datasets and domains demonstrate that our framework significantly enhances the performance of sequential recommendation models on the ZCDSR task. By addressing domain bias and improving the transfer of sequential patterns, our method offers a scalable and robust solution for better knowledge transfer, enabling improved zero-shot recommendations across domains.
\end{abstract}



\maketitle

\vspace{-5pt}
\section{Introduction}
\begin{table*}[ht]
\centering
\small
\caption{Illustration of domain semantic bias in item descriptions from Industrial \& Scientific (IS) vs. Video Games (VG).}

\label{tab:text_domain}
\begin{tabular}{p{3.5cm}p{5.5cm}p{5.5cm}}
\toprule
\textbf{Aspect}         & \textbf{Industrial \& Scientific (IS)}                                         & \textbf{Video Games (VG)} \\ 
\midrule
\textbf{Item Title}      & \textbf{XXX Magnetics Magnet Fishing Kit}                                       & \textbf{XXX Controller Faceplate} \\ 
\textbf{Example Description} & 
"Includes powerful rare earth magnet  with double braided polyester rope and carabiner. This strong magnet is ideal for ocean piers, lake docks, and bridges. Great for salvage and treasure hunting." &
"Ultra fits for Xbox One X \& One S controller; Completely fits flush on all sides; The shadow purple color looks great with a smooth grip, anti-slip, and sweat-free for long periods of gameplay." \\ 
\textbf{Key Vocabulary}      & \textit{magnet fishing, rare earth magnet, pull force, threadlocker, paracord}       & \textit{faceplate, side rails, anti-slip, soft touch, screwdriver}          \\ 
\textbf{Content Focus}   & Focused on durability, kit completeness, and outdoor usability, such as strength for fishing and treasure hunting.      & Focused on controller modification, grip quality, and aesthetic customization for enhanced gaming experience.  \\ 
\bottomrule
\end{tabular}

\end{table*}

Zero-shot cross-domain sequential recommendation (ZCDSR) extends beyond conventional recommendation tasks~\cite{rendle2012bpr,kang2018self} by addressing the critical challenge of making accurate predictions in unseen domains where no prior interaction data exists. This capability is indispensable in dynamic environments—such as new markets, product categories, or emerging user segments—where collecting domain-specific interaction data is often impractical, and it is particularly relevant in modern applications like e-commerce, streaming platforms, and online education, which frequently encounter rapid changes and the introduction of new domains. Traditional models, which rely heavily on domain-specific training data, often struggle to adapt in such scenarios, resulting in poor recommendation quality and a diminished user experience. ZCDSR plays a crucial role in overcoming this limitation by enabling recommendation systems to make effective predictions without domain-specific data, ensuring they remain robust and relevant in diverse and evolving contexts. ZCDSR represents a more challenging setting compare to few-shot adaptation, as it requires the system to generalize effectively without any prior exposure to the target domain, fundamentally challenging the model’s generalizability.

 Existing works in zero-shot recommendation~\cite{ding2021zero, feng2021zero, wang2023pre, he2023large} leverage metadata—such as text, images, and popularity—to transfer knowledge from source to unseen domains. The emergence of large language models (LLMs) has further advanced this direction by capturing rich item semantics and user intent through pretrained textual representations. Current LLM-based methods (LLM4Rec) can be categorized into: (1) \textit{direct recommenders}\cite{yue2023llamarec, bao2023tallrec, ji2024genrec}, which generate predictions directly from textual inputs, and (2) \textit{feature encoders}\cite{qiu2021u, rajput2023recommender, li2023exploring, li2024multi}, which generate semantic embeddings for downstream recommendation models.

However, when used as feature encoders, LLMs face a key challenge: \textbf{\textit{domain semantic bias}}. This arises from mismatched vocabularies and content focus across domains, hindering the transferability of LLM-generated embeddings. For instance, as shown in~\Cref{tab:text_domain}, the Industrial \& Scientific domain emphasizes technical attributes like \textit{magnet strength} and \textit{durability}, catering to \textit{utility-focused users}, whereas Video Games focus on \textit{visual design} and \textit{grip comfort}, appealing to \textit{gamers}. When embeddings trained in one domain are applied to another, the misaligned semantics can degrade recommendation performance. This highlights the need for methods that explicitly address cross-domain semantic misalignment to fully realize the zero-shot potential of LLMs.
 
 This divergence in semantic focus highlights the domain-specific nature of item descriptions, which can significantly impact LLM-generated embeddings. When item embeddings trained on one domain are applied to another, the model may produce suboptimal recommendations due to the misalignment of feature representations. For instance, an LLM trained on the technical, performance-oriented descriptions of the Industrial \& Scientific domain may fail to capture the subjective, experiential attributes emphasized in the Video Games domain. Such misalignment underscores the need for improved methods to bridge semantic gaps across domains, ensuring more robust and generalizable recommendation systems.
 
\textbf{Our Insight:}  
Domain-specific semantic bias exists in LLM-based zero-shot sequential recommenders. Therefore, it is crucial to strike a balance between generalization across domains and the preservation of domain-specific characteristics. Specifically, aligning item embeddings across domains while retaining the unique attributes of each domain is key to achieving effective cross-domain recommendation. Moreover, user behaviors tend to exhibit similarities across domains~\cite{ pareto1897cursus, Wang_2024}, particularly in terms of the temporal or relational structures that govern item progression within a sequence, such as dynamic relationships and preference transitions. By capturing and leveraging these sequential patterns, we can enhance the recommendation process and make more accurate predictions, even in the absence of target-domain interaction data.

\textbf{Present Work:}  
We propose a novel model-agnostic \textbf{LLM}-based \textbf{Rec}ommendation \textbf{G}eneralization framework for domain semantic bias, \textbf{LLM-RecG}. LLM-RecG comprehensively addresses domain bias by capturing transferable sequential patterns while preserving domain-specific nuances, (\eg distinct vocabularies, interaction behaviors, and content focuses), ensuring accurate recommendations in unseen target domains.
Specifically, we introduce  a novel training objective that balances \textit{inter-domain compactness} and \textit{intra-domain diversity} at the \textit{item level}. \textit{Inter-domain compactness} ensures that item embeddings are closely aligned across different domains, facilitating knowledge transfer and reducing domain-specific biases. In contrast, \textit{intra-domain diversity} maintains the fine-grained distinctions among items within the same domain, ensuring the model does not overfit to dominant source-domain features." These two complementary objectives balance cross-domain alignment and domain-specific nuance, resulting in improved generalization and more accurate recommendations across domains. Furthermore, we transfer sequential patterns from the source domain to the target domain. Unlike item-level embedding generalization, sequential patterns capture how items are interacted with in specific sequences, providing critical context for user preferences. By clustering source user sequences into patterns, LLM-RecG learns to aggregate information from relevant sequential patterns through attention mechanisms during target domain inference. This enables dynamic adaptation of user embeddings without requiring target-domain interaction data, ensuring effective zero-shot recommendations.
In summary, our main contributions are as follows:

\begin{description}[labelindent=5pt, labelsep=0pt, topsep=0pt, leftmargin=!, font=\normalfont\bfseries]
\item \textbf{Domain Semantic Bias-aware Framework: }To the best of our knowledge, we are the \textbf{first} to identify and address the issue of domain semantic bias in LLM-based zero-shot cross-domain sequential recommendation (ZCDSR). Previous work typically focuses on leveraging domain-agnostic representations and overlooks the domain semantic bias, which significantly impact the transfer process. We, on the other hand, lay the groundwork for more effective knowledge transfer across domains by analyzing the sources of domain semantic bias and highlighting its effect on the performance of zero-shot recommendation tasks, which in turn leads to more accurate and reliable zero-shot predictions in unseen domains. Our results, both qualitative and quantitative, demonstrate that addressing this bias is crucial for ZCDSR and significantly enhances the model’s generalizability.
    
\item \textbf{Dual-level Generalization: }We introduce a new generalization approach that enhances both \textit{item}-level and \textit{sequence}-level generalization. In contrast to prior methods that primarily focus on static item embeddings or sequential patterns individually, these approaches often fail to holistically address both levels. Our approach combines item-level generalization via inter-domain compactness and intra-domain diversity with sequential pattern transfer across domains using attention mechanisms. This dual-level generalization results in more effective zero-shot recommendations, particularly in sparse or unseen domains, without relying on target-domain interaction data. Through extensive experiments, we show that LLM-RecG robust performance across domains, even with sparse and unseen target domains.
    
\end{description}
\vspace{-10pt}  
\section{Problem Definition}



In this section, we formally define the \textit{zero-shot cross-domain sequential recommendation} (ZCDSR) task and introduce the notations used throughout this work.

Let \( \mathcal{D}_s = \{\mathcal{V}_s, \mathcal{U}_s, \mathcal{X}_s \} \) represent the source domain, where \( \mathcal{V}_s = \{v^s_1, v^s_2, \dots, v^s_{|\mathcal{V}_s|}\} \) denotes the set of items, and \( \mathcal{U}_s = \{u^s_1, u^s_2, \dots, u^s_{|\mathcal{U}_s|}\} \) is the set of users in domain \( \mathcal{D}_s \). Each item \( v_i^s \) is associated with metadata \( x_i^s \), forming \( \mathcal{X}_s \), the metadata set mapped one-to-one to the items. The goal of sequential recommendation in the source domain is to learn a scoring function that predicts the next item \( v_{j, t}^s \) for a user \( u_j^s \), given their interaction history \( \mathcal{H}_j = \{v_{j, 0}^s, v_{j, 1}^s, \dots, v_{j, t-1}^s \} \). Formally, the scoring function is defined as:
$\mathcal{F} (v_{j, t}^s \mid \mathcal{H}_j, \mathcal{X}_s)$,
where \( \mathcal{F} \) outputs a ranking score for candidate items based on the user’s historical interactions and the item metadata.

\textbf{Zero-shot Cross-Domain Sequential Recommendation: } Given a new target domain \( \mathcal{D}_t = \{\mathcal{V}_t, \mathcal{U}_t, \mathcal{X}_t\} \), where \( \mathcal{V}_t \cap \mathcal{V}_s = \emptyset \) and \( \mathcal{U}_t \cap \mathcal{U}_s = \emptyset \), the objective is to produce a scoring function \( \mathcal{F}' \) for \( \mathcal{D}_t \) without training on \( \mathcal{D}_t \) directly. This setting assumes no overlap in items or users between the source and target domains, making it a purely zero-shot transfer scenario.

We specifically focus on the zero-shot transfer setting rather than a few-shot setting, as it presents a more challenging yet crucial problem to study. Addressing this challenge is essential and fundamental for improving the scalability and adaptability of sequential recommendation systems across diverse domains.

\begin{figure*}[htb]
    \centering

    \includegraphics[width=0.95\linewidth]{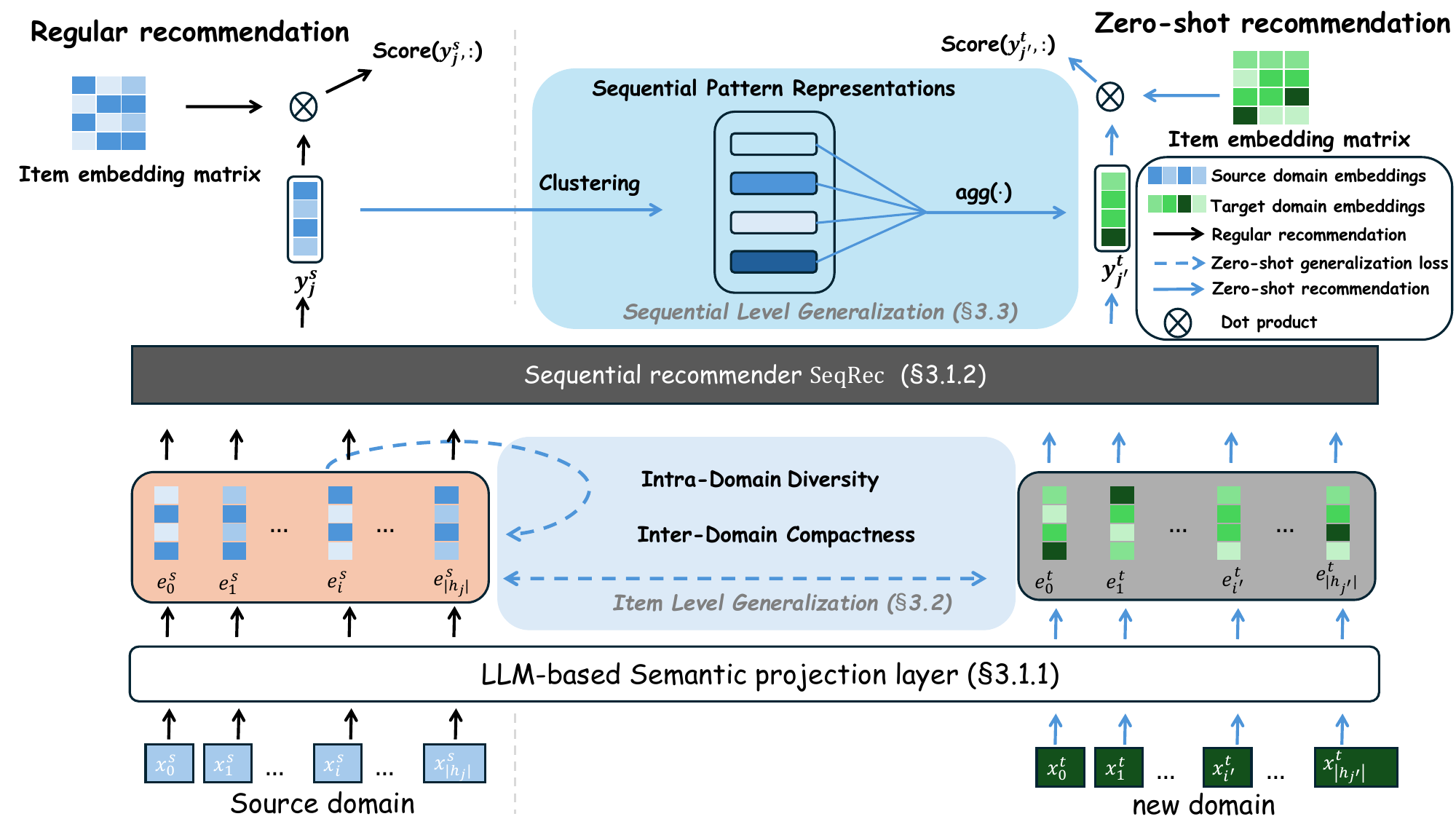}
    \caption{The model framework of LLM-RecG.} 
    \label{fig:framework}
\end{figure*}

\vspace{-5pt}

\section{Methodology}
\label{sec:method}
\vspace{-2pt}
In this section, we present LLM-RecG, a model-agnostic generalization framework designed to address domain semantic bias in zero-shot cross-domain sequential recommendation (ZSCDSR). Unlike few-shot adaptation, ZSCDSR requires models to generalize to entirely unseen domains without any target-domain interactions. LLM-RecG tackles this challenge by enhancing the scalability and adaptability of sequential recommenders, enabling robust performance without further tuning.

At its core, LLM-RecG is built upon a semantic sequential framework (\cref{sec:base_method}), which generalizes existing sequential recommenders to ZSCDSR by incorporating an LLM-based semantic projection layer (\cref{sec:semantic_proj}). To further strengthen the generalization process, we propose a dual-level generalization strategy (\cref{sec:item-level} and~\cref{sec:seq-level}), which mitigates domain semantic discrepancies and improves the generalizability of the model in zero-shot scenarios.
\vspace{-5pt}
\subsection{Semantic Sequential Framework}
\label{sec:base_method}
The semantic sequential framework forms the backbone of LLM-RecG. It encodes item metadata into rich, high-dimensional embeddings and then mapping these embeddings into a space that captures sequential dependencies between items using any exisiting sequential recommenders, making it adaptable across domains. The use of an LLM-based semantic projection layer (\cref{sec:semantic_proj}) ensures that the model can process and integrate item-specific features in a way that captures the latent relationships between items in a sequence, allowing us to not only retain the domain semantic bias but also make it transferable to a target domain in a zero-shot setup.
\subsubsection{LLM-based semantic projection layer.} 
\label{sec:semantic_proj}
We leverage the power of LLMs to capture rich semantic information from item metadata. Specifically, we employ LLMs to extract semantic embeddings that encode contextual information from textual descriptions. For each item \( v^s_i \in \mathcal{V}_s \) in the source domain, we aggregate its associated metadata (\eg, title, features, and description) into a unified textual description \( x^s_i \), following a predefined template.
We then adopt $\mathcal{E}$, a LLM-based semantic encoder to generate a high-dimensional semantic embedding $\mathbf{e}^{\text{sem}}_i$ of the item $v^s_i$:
\begin{equation}
\mathbf{e}^{\text{sem}}_i = \mathcal{E}(\text{x}^s_i),
\end{equation}
where \( \mathbf{e}^{\text{sem}}_i \in \mathbb{R}^{d_h} \) and \( d_h \) denotes the dimensionality of the semantic embedding space. Then, we apply a projection layer that maps the embeddings into a lower-dimensional latent space to reduce dimensionality and align the embeddings with user interaction patterns, enhancing the model's ability to capture sequential dependencies:
\begin{equation}
\mathbf{e}^s_i = \mathbf{W}_p \mathbf{e}^{\text{sem}}_i + \mathbf{b}_p,
\end{equation}
where \( \mathbf{e}^s_i \in \mathbb{R}^{d_l} \), \( \mathbf{W}_p \in \mathbb{R}^{d_l \times d_h} \), and \( \mathbf{b}_p \in \mathbb{R}^{d_l} \) are learnable parameters of the projection layer, with \( d_l < d_h \).


\subsubsection{Sequential Dependencies Modeling.}
\label{sec:sequential}
Once the semantic embeddings for the items are obtained, we turn to sequential recommenders to capture the inherent sequential dependencies between items in a user's interaction history.
We denote the sequence of low-dimensional embeddings corresponding to user $u^s_j$'s interaction history \( \mathcal{H}_j^s = \{v_{j, 0}^s, v_{j, 1}^s, \dots, v_{j, |\mathcal{H}_j|}^s \}\) in the source domain as:
\begin{equation}
\mathbf{H}^s_j = \{\mathbf{e}_{j, 0}^s, \mathbf{e}_{j, 1}^s, \dots, \mathbf{e}_{j, |\mathcal{H}_j|}^s\},
\end{equation}
where \( \mathbf{H}^s_j \in \mathbb{R}^{|h^s_j| \times d_l} \) is the sequence of embeddings for user \( u^s_j \) in the source domain. This sequence is then passed to any sequential recommendation models \( \mathit{SeqRec} \), forming the core of our model-agnostic approach. These models process the sequence and output a user representation: 
\begin{equation}
\mathbf{y}^s_j = \text{SeqRec}(\mathbf{H}^s_j),
\end{equation}
where \( \mathbf{y}^s_j \in \mathbb{R}^{d_l} \) encodes the user’s preferences based on their interaction history in the source domain.

To predict the relevance of candidate items, we compute a score function as follows:
\begin{equation}
\text{score}(h^s_j, v^s_i) = <{\mathbf{y}^s_j},\mathbf{e}^s_i>
\end{equation}
where \( \mathbf{e}^s_i \in \mathbb{R}^{d_l} \) is the embedding of the candidate item \( v^s_i \) in the source domain and $<\cdot>$ denotes the dot product. The resulting score is then used to rank candidate items for recommendation.

To optimize the model, we adopt the Bayesian Personalized Ranking (BPR) loss~\cite{rendle2012bpr}, which maximizes the pairwise ranking between positive and negative items. For a given user \( u^s_j \) in the source domain, the loss is defined as:
\begin{equation}
\mathcal{L}_{\text{rec}} = -\sum_{j} \sum_{(i^+, i^-)} \log \sigma\left(\text{score}(u^s_j, v^s_{i^+}) - \text{score}(u^s_j, v^s_{i^-})\right),
\end{equation}
where \( i^+ \) and \( i^- \) are positive and negative items, respectively, and \( \sigma(\cdot) \) is the sigmoid function.

When performing zero-shot inference on a new domain, the same pretrained $\mathcal{E}$ and projection layers (\ie $\mathbf{W}_p$ and  $ \mathbf{b}_p$) are used to process new items. For each item \( v_{i'}^t \in \mathcal{V}^t \) in the target domain, the textual attributes \( x_{i'} \) are transformed into semantic embeddings and subsequently projected into the low-dimensional space:
\begin{equation}
\mathbf{e}_{i'}^t = \mathbf{W}_p \mathbf{e}_{i'}^{\text{sem}} + \mathbf{b}_p.
\end{equation}
This ensures that new item embeddings in the target domain are seamlessly integrated into the scoring framework without requiring model fine-tuning.


\vspace{-5pt}
\subsection{Item-Level Generalization}
\label{sec:item-level}
At the item level, we propose that semantic embeddings of items should satisfy the following two key properties:

\begin{definition}[Inter-Domain Compactness] Inter-domain compactness ensures that item embeddings from different domains are closely aligned in the embedding space, reducing domain semantic biases and enabling effective knowledge transfer. \end{definition}

\begin{definition}[Intra-Domain Diversity] Intra-domain diversity ensures that item embeddings within the same domain remain distinct, capturing fine-grained variability and preventing representation collapse. \end{definition}
\subsubsection{Inter-Domain Compactness}

As defined, inter-domain compactness aligns item embeddings from different domains by minimizing the entropy of the similarity distribution between embeddings and the centers of other domains. Let \(\{\mathbf{c}_d | d \in \{\mathcal{D}_s, \mathcal{D}_t\}\}\) represent the domain centers, and \(\{\mathbf{e}^s_i, \mathbf{e}^t_i | v^s_i \in \mathcal{V}_s, v^t_i \in \mathcal{V}_t\}\) denote the item embeddings from the source and target domains.

The domain center \( \mathbf{c}_d \) for a domain \( d \) (\(d \in \{\mathcal{D}_s, \mathcal{D}_t\}\)) is defined as the mean of all embeddings belonging to that domain:
\begin{equation}
\mathbf{c}_d = \frac{1}{|\mathcal{V}_d|} \sum_{v_i^d \in \mathcal{V}_d} \mathbf{e}^d_i,
\end{equation}
where \( \mathcal{V}_d \) represents the set of items in domain \( d \), and \( |\mathcal{V}_d| \) is the number of items in that domain.

To focus specifically on inter-domain alignment, the inter-domain compactness loss is formulated as:
\begin{equation}
\mathcal{L}_{\text{inter}} = \sum_{v_i \in \mathcal{V}} \sum_{\substack{d \in \{\mathcal{D}_s, \mathcal{D}_t\} \\ d \neq d_{i}}} Q_{id} \log Q_{id},
\end{equation}
where \( d_{i} \) indicates the domain of item \( v_i \), and \( Q_{id} \) is the probability of embedding \( \mathbf{e}^d_i \) being associated with domain \( d \). This probability is computed as:
\begin{equation}
Q_{id} = \frac{\exp\left(\text{cos}(\mathbf{e}^d_i, \mathbf{c}_d) / \tau\right)}{\sum_{\substack{d' \in \{\mathcal{D}_s, \mathcal{D}_t\} \\ d' \neq d_{i}}} \exp\left(\text{cos}(\mathbf{e}^d_i, \mathbf{c}_{d'}) / \tau\right)},
\end{equation}
where \( \text{cos}(\cdot) \) represents cosine similarity, and \( \tau > 0 \) is a temperature parameter that controls the sharpness of the similarity distribution.

Minimizing \( \mathcal{L}_{\text{inter}} \) encourages embeddings to align with the centers of other domains, enhancing inter-domain compactness and reducing domain semantic biases. However, it alone is insufficient to ensure diversity within the same domain. To address this, we propose an additional objective for \textit{intra-domain diversity}, which prevents embeddings from collapsing into overly similar representations and preserves fine-grained variability within each domain.

\subsubsection{Intra-Domain Diversity}

As defined, intra-domain diversity ensures distinctiveness among embeddings within the same domain, effectively capturing item-level variability and preventing representation collapse. For a domain \( d \in \{\mathcal{D}_s, \mathcal{D}_t\} \) with item embeddings \(\{\mathbf{e}^d_i | v^d_i \in \mathcal{V}_d\}\), the intra-domain diversity loss is formulated as:
\begin{equation}
\mathcal{L}_{\text{intra}} = - \sum_{d \in \{\mathcal{D}_s, \mathcal{D}_t\}} \frac{1}{|\mathcal{V}_d|} \sum_{v^d_i \in \mathcal{V}_d} \sum_{v^d_j \in \mathcal{V}_d} P_{ij} \log P_{ij},
\end{equation}

where \( P_{ij} \) denotes the similarity-based probability between items \( v^d_i \) and \( v^d_j \) within domain \( d \). This probability is computed using cosine similarity with temperature scaling:
\begin{equation}
P_{ij} = \frac{\exp\left(\text{cos}(\mathbf{e}^d_i, \mathbf{e}^d_j) / \tau\right)}{\sum_{v^d_k \in \mathcal{V}_d} \exp\left(\text{cos}(\mathbf{e}^d_i, \mathbf{e}^d_k) / \tau\right)},
\end{equation}
where \( \tau > 0 \) regulates the sharpness of the similarity distribution. Lower \( \tau \) sharpens the distribution, amplifying differences, while higher \( \tau \) smooths it, promoting more uniform probabilities. Maximizing this entropy-based objective discourages embeddings within the same domain from collapsing into similar points, fostering richer intra-domain representation.

The generalization loss combines intra-domain diversity and inter-domain compactness:
\begin{equation}
\mathcal{L}_{\text{gen}} = -\alpha \mathcal{L}_{\text{intra}} + \beta \mathcal{L}_{\text{inter}},
\end{equation}
where \(\alpha > 0\) and \(\beta > 0\) control the balance between the two terms. This formulation promotes variability within domains while aligning embeddings across different domains, mitigating domain semantic biases. To balance the relative influence, \(\beta\) is scaled relative to \(\alpha\) based on the number of domains \(|\{\mathcal{D}_s, \mathcal{D}_t\}|\) and the total number of items \(|N|\):
\begin{equation}
\beta = \alpha \frac{|N|}{|\{\mathcal{D}_s, \mathcal{D}_t\}|^3}.
\end{equation}
This scaling ensures that inter-domain compactness contributes appropriately without overpowering intra-domain diversity.

To reduce computational complexity, we adopt a sampling strategy that approximates intra-domain and inter-domain components using batch data, significantly lowering overhead while enabling efficient integration with the recommendation loss.

The generalization loss \(\mathcal{L}_{\text{gen}}\) complements the primary recommendation objective \(\mathcal{L}_{\text{rec}}\) during training. The overall objective function is:
\begin{equation}
\mathcal{L}_{\text{total}} = \mathcal{L}_{\text{rec}} + \mathcal{L}_{\text{gen}}.
\end{equation}

To preserve semantic information and enhance generalization, we introduce an additional projection layer. This layer maps semantic embeddings into another latent space, complementing the initial projection. The final item embeddings are obtained by merging the outputs of both projection layers.
\vspace{-5pt}
\subsection{Sequence-level Generalization}
\label{sec:seq-level}
Item-level generalization focuses on aligning static item embeddings across domains, but it overlooks the sequential dependencies in user behavior, which are crucial for understanding preferences and predicting future interactions. A sequential pattern abstracts the temporal or relational structure governing item progression within a sequence, capturing dynamic relationships and transitions in user preferences. These patterns highlight commonalities in user behavior across domains, making them particularly valuable for knowledge transfer. To bridge the gap left by item-level generalization, we propose leveraging these sequential patterns from the source domain to enhance user sequence representations during target domain inference.

\subsubsection{Sequential Pattern Extraction from Source Domain}
We extract \textit{sequential patterns} that encapsulate common user behavioral trajectories by clustering the sequence embeddings of users from the source domain. Let \( \mathbf{y}^s_j \in \mathbb{R}^{d_l} \) denote the sequence embedding of user \( u^s_j \) in the source domain, obtained by encoding the interaction history \( h^s_j = \{v^s_1, v^s_2, \dots, v^s_{|h^s_j|}\} \).

The set of \( k \) sequential patterns \( S = \{ s_1, s_2, \dots, s_k \} \) is extracted by applying \( k \)-means clustering~\cite{macqueen1967some} over the sequence embeddings:
\begin{equation}
S = \text{k-means}(\{ \mathbf{y}^s_j \ | \ u^s_j \in \mathcal{U}_s \}),
\end{equation}
where \( \mathcal{U}_s \) represents the set of users in the source domain. Each pattern \( s_i \in \mathbb{R}^{d_l} \) corresponds to the centroid of a cluster, representing shared sequential behaviors observed in the source domain.

\subsubsection{Soft Sequential Pattern Attention for Target Sequences}
For zero-shot inference in the target domain, target user sequences are encoded to produce sequence embeddings \( \mathbf{y}^t_j \) for user \( u^t_j \), based on their interaction history \( h^t_j = \{v^t_1, v^t_2, \dots, v^t_{|h^t_j|}\} \). Since target domain interaction data is unavailable during training, the model leverages source domain sequential patterns to guide the recommendation process. Rather than selecting the closest pattern directly, a soft attention mechanism is applied to aggregate information from multiple patterns.

The similarity between the target sequence embedding \( \mathbf{y}^t_j \) and each sequential pattern \( \mathbf{s}_i \) is computed using cosine similarity:
\begin{equation}
s^t_{j,i} = \text{cos}(\mathbf{y}^t_j, \mathbf{s}_i) = \frac{\mathbf{y}^t_j \cdot \mathbf{s}_i}{\| \mathbf{y}^t_j \| \| \mathbf{s}_i \|}.
\end{equation}

These similarity scores are normalized using the softmax function to obtain attention weights over the sequential patterns:
\begin{equation}
\alpha^t_{j,i} = \frac{\exp(s^t_{j,i})}{\sum_{l=1}^{k} \exp(s^t_{j,l})}.
\end{equation}

The attended sequential pattern representation \( \tilde{s}^t_j \) for user \( u^t_j \) is computed as the weighted sum of the patterns:
\begin{equation}
\tilde{s}^t_j = \sum_{i=1}^{k} \alpha^t_{j,i} \mathbf{s}_i.
\end{equation}

\subsubsection{Fusion of User Embedding and Pattern Representation}
To integrate source domain patterns while preserving target-specific information, the target user embedding \( \mathbf{y}^t_j \) is concatenated with the attended pattern representation \( \tilde{s}^t_j \):
\begin{equation}
\mathbf{f}^t_j = [\mathbf{y}^t_j ; \tilde{\mathbf{s}}^t_j],
\end{equation}
where \( [\cdot ; \cdot] \) denotes concatenation. The fused representation \( \mathbf{f}^t_j \in \mathbb{R}^{2d_l} \) is then projected back into the original embedding space:
\begin{equation}
\mathbf{g}^t_j = \mathbf{W}_f \mathbf{f}^t_j,
\end{equation}
where \( \mathbf{W}_f \in \mathbb{R}^{d_l \times 2d_l} \) is a learnable projection matrix. The resulting fused embedding \( \mathbf{g}^t_j \) captures both the user's target-specific preferences and the transferable sequential patterns from the source domain. It is subsequently used for next-item prediction.

\begin{table}[t]
\centering
\caption{Processed Dataset Statistics.}
\vspace{-5pt}
\label{tab:dataset_stats}
\scalebox{0.9}{
\begin{tabular}{lrrc}
\toprule
\textbf{Dataset} & \textbf{Items} & \textbf{Users} & \textbf{Avg. Seq. Len.} \\
\midrule
Industrial \& Scientific (IS) & 24,315 & 47,905 & 8.04 \\
Video Games (VG) & 24,030 & 89,746 & 8.60 \\
Musical Instruments (MI) & 23,618 & 55,787 & 8.88 \\
Steam & 8,156 & 26,541 & 19.68 \\
\bottomrule
\end{tabular}
}
\vspace{-10pt}
\end{table}

\begin{table*}[t]
\centering
\renewcommand{\arraystretch}{1.0}
\setlength{\tabcolsep}{4pt}
\caption{
Zero-shot performance across all source–target domain pairs.Throughout all result tables, we report Recall@10 (R@10) and NDCG@10 (N@10) in percentage (\%), unless otherwise specified. Base models include two variants: \texttt{-Sem} (with LLM-based semantic embeddings) and \texttt{-RecG} (with the proposed generalization loss). 
Our method consistently outperforms all semantic-only variants across diverse domain pairs. It enhances robustness to domain shifts, reducing variability in performance across source–target combinations. Notably, even in challenging cross-platform transfers from Steam, \texttt{-RecG} delivers substantial gains (e.g., BERT4Rec-RecG improves N@10 by 28.9\% over its \texttt{-Sem} counterpart when transferring to IS), effectively mitigating domain semantic bias. \vspace{-5pt}
}

\label{tab:zeroshot_performance}
\scalebox{0.85}{
\begin{threeparttable}
\begin{tabular}{cccccccccccccccccc}
\toprule
\textbf{Target} & \textbf{Source} & 
\multicolumn{2}{c}{GRU4Rec-Sem} & 
\multicolumn{2}{c}{SASRec-Sem} & 
\multicolumn{2}{c}{BERT4Rec-Sem} & 
\multicolumn{2}{c}{UniSRec} & 
\multicolumn{2}{c}{RecFormer} & 
\multicolumn{2}{c}{GRU4Rec-RecG} & 
\multicolumn{2}{c}{SASRec-RecG} & 
\multicolumn{2}{c}{BERT4Rec-RecG} \\
& & R@10 & N@10 & R@10 & N@10 & R@10 & N@10 & R@10 & N@10 & R@10 & N@10 & R@10 & N@10 & R@10 & N@10 & R@10 & N@10 \\
\midrule
\multirow{3}{*}{IS} 
& VG    & 20.02 & 9.69 & 30.27 & 16.51 & 27.22 & 14.57 & 27.29 & 14.62 & \underline{31.38} & \underline{17.96} & 27.94 & 15.47 & \textbf{36.94*} & \textbf{18.32*} & 35.23 & 17.67 \\ 
& MI    & 21.78 & 10.48 & 29.88 & 16.68 & 29.94 & 16.31 & 29.67 & 16.12 & \underline{32.95} & \underline{18.43} & 28.97 & 16.39 & \textbf{38.59*} & \textbf{20.19*} & 37.21 & 20.10 \\
& Steam & 15.92 & 7.95 & \underline{23.22} & \underline{12.94} & 22.91 & 12.86 & 22.18 & 13.26 & 23.16 & 13.74 & 21.94 & 12.77 & 27.46 & 16.43 & \textbf{29.15*} & \textbf{16.82*} \\
\midrule
\multirow{3}{*}{VG} 
& IS    & 17.01 & 8.13 & \underline{32.97} & \underline{18.68} & 31.84 & 17.78 & 30.91 & 18.14 & 31.54 & 18.70 & 27.66 & 15.58 & \textbf{37.10*} & \textbf{22.59*} & 35.81 & 21.53 \\
& MI    & 23.30 & 11.61 & \underline{34.73} & \underline{19.46} & 32.94 & 18.14 & 34.62 & 19.95 & 36.67 & 21.09 & 30.84 & 16.34 & \textbf{41.59*} & \textbf{23.67*} & 40.58 & 22.27 \\
& Steam & 16.19 & 7.74 & \underline{27.39} & \underline{16.33} & 26.17 & 15.52 & 26.49 & 14.85 & 27.54 & 16.74 & 23.73 & 12.17 & 29.46 & 17.49 & \textbf{29.26*} & \textbf{18.34*} \\
\midrule
\multirow{3}{*}{MI} 
& IS    & 16.35 & 7.81 & \underline{26.20} & \underline{14.11} & 24.82 & 13.28 & 25.16 & 13.85 & 27.69 & 17.29 & 23.89 & 12.54 & \textbf{32.16*} & \textbf{18.24*} & 30.10 & 17.06 \\
& VG    & 20.08 & 9.75 & \underline{27.74} & \underline{15.06} & 25.13 & 13.32 & 26.91 & 15.62 & 28.44 & 17.67 & 26.83 & 14.18 & \textbf{34.42*} & \textbf{19.47*} & 32.51 & 18.81 \\
& Steam & 14.70 & 6.29 & \underline{20.46} & \underline{11.83} & 20.29 & 11.73 & 21.75 & 12.38 & 22.39 & 13.54 & 20.75 & 11.05 & 25.84 & 13.49 & \textbf{26.74*} & \textbf{13.92*} \\
\midrule
\multirow{3}{*}{Steam} 
& IS    & 16.24 & 7.62 & \underline{21.54} & \underline{11.28} & 19.75 & 9.97 & 20.36 & 11.36 & 21.18 & 11.83 & 21.44 & 12.37 & \textbf{23.59*} & \textbf{12.48*} & 22.49 & 12.85 \\
& VG    & 18.52 & 7.80 & \underline{25.72} & \underline{13.45} & 20.53 & 11.62 & 21.52 & 11.74 & 23.06 & 12.47 & 22.61 & 12.59 & \textbf{29.26*} & \textbf{14.93*} & 25.11 & 13.79 \\
& MI    & 17.73 & 7.75 & \underline{22.74} & \underline{11.36} & 19.13 & 9.41 & 21.86 & 12.01 & 23.29 & 12.31 & 22.87 & 12.14 & \textbf{26.13*} & \textbf{13.71*} & 22.72 & 12.97 \\
\bottomrule
\end{tabular}
\begin{tablenotes}
\footnotesize
\item Throughout all result tables, the best results are marked with \textit{*}($t$-test, \textit{p} $\leq$ 0.05) compared with the best \underline{baselines}, unless otherwise specified.
\end{tablenotes}
\end{threeparttable}
}
\vspace{-5pt}
\end{table*}

\section{Experiments}

We conduct extensive experiments on three real-world datasets to evaluate the performance of LLM-RecG, following the settings outlined in ~\Cref{sec:method}. To guide our analysis, we address the following research questions (RQs):  
\begin{itemize}[left=0pt]
    \item \textbf{RQ1}: How well does LLM-RecG enhance baseline models in both in-domain and zero-shot scenarios?
    \item \textbf{RQ2}: What is the effectiveness of each component in LLM-RecG?
    \item \textbf{RQ3}: How does the key parameter impact performance?
    \item \textbf{RQ4}: How does LLM-RecG affect the alignment and uniformity of item embeddings across domains?
\end{itemize}

\subsection{Experimental Setup}
\subsubsection{Datasets}
We select three subsets i.e., \textit{Video Games, Industrial \& Scientific, Musical Instruments} from the Amazon Review dataset~\cite{hou2024bridging} with another cross-platform Steam Dataset~\cite{kang2018self}, which contain user-item interactions and textual metadata for all items. Following the preprocessing outlined in previous works~\cite{rendle2010factorizing, he2016fusing, kang2018self, li2021extracting}, we filter out all users and items with fewer than ten interactions. Additionally, we exclude items that lack textual features, such as descriptions or attributes. For all datasets, the last item in each sequence is used for testing, while the penultimate item is used for validation. During evaluation, we adopt a negative sampling strategy, where the ground truth item is ranked against 100 randomly sampled negative items instead of the entire item set. The statistics of the processed datasets are summarized in Table~\ref{tab:dataset_stats}.

\subsubsection{Baseline \& Evaluation Metrics}
We present our method as a foundational framework and validate its effectiveness through comprehensive evaluations on three widely adopted recommendation models: \textbf{GRU4Rec}~\cite{hidasi2015session}, \textbf{SASRec}~\cite{kang2018self}, and \textbf{Bert4Rec}~\cite{sun2019bert4rec}. We denote variants enhanced with LLM-based semantic embeddings as \texttt{-Sem} and those further equipped with our proposed generalization loss as \texttt{-RecG}. Additionally, we also compared with two state-of-the-art text-only cross-domain recommendation methods, \textbf{UniSRec}~\cite{hou2022towards} and \textbf{RecFormer}~\cite{li2023text}.
For performance assessment, we adopt two widely used metrics: Recall \textbf{R@k} and Normalized Discounted Cumulative Gain (\textbf{N@k}).

\subsubsection{Implementation Details}
Our models are implemented using Python 3.9.20 and PyTorch 2.5.1~\footnote{ Code and data available at: \url{https://github.com/yunzhel2/LLM-RecG.}}. For optimization, we use the Adam optimizer across all models and adopt mini-batches with a batch size of 128. For LLM-based semantic encoder $\mathcal{E}$ mentioned in~\Cref{sec:base_method}, We adopt \textbf{LLM2Vec}~\cite{llm2vec} with Llama 3-8B~\cite{dubey2024llama}.To ensure fair comparisons, we perform a grid search to identify the optimal hyperparameters for each model. The search space includes embedding dimensionalities \{64, 128, 256\}, the number of layers \{1, 2, 3\}, dropout rates \{0.2, 0.3, 0.5\}, learning rates \{0.001, 0.0005, 0.0001\}, and, where applicable, the number of attention heads \{1, 2, 4\}, $\alpha \in \{0.05, 0.01,,0.005,0.001,0.0005,0.0001\}$ . For evaluation, we use $k\in \{5,10,20\}$ for both R@k and N@k. Due to the page limitation, only $k=10$ is reported. For all models, we repeat experiments 5 times with different random seeds and report the average performance.

\subsection{Overall Comparison(RQ1)}

\subsubsection{ZCDSR performance.}
For the ZCDSR task, we follow the experimental settings outlined in Section~\ref{sec:base_method}. From Table~\ref{tab:zeroshot_performance}, we observe the following key insights: a) Across all models and domain pairs, our method achieves substantial improvements over variants that directly use LLM-based semantic embeddings, with average gains exceeding 10\% compared with best baselines even when transfering cross-platform between Amazon and Steam. This demonstrates the effectiveness of the proposed generalization framework in enhancing zero-shot recommendation performance while addressing domain semantic bias. 
b) Our method also improves robustness to domain shifts. For instance, GRU4Rec-Sem shows significant performance degradation depending on the relationship between source and target domains, with performance dropping by over 6\% in R@10 in the VG domain when sourced from IS or MI. In contrast, our approach mitigates this variability, delivering more consistent and reliable results across diverse domain pairs.
c)  We find that models trained on the Steam dataset generally perform worst when transferred to other domains, likely due to semantic discrepancies in item descriptions. For example, BERT4Rec-Sem achieves only 9.97\% at N@10 when transferring from Steam to IS—much lower than transfers from MI 17.06\% or VG 17.67\%. Despite this, our \texttt{-RecG} variants significantly improve performance under such domain shifts. In the same setting, BERT4Rec-RecG by relatively 28.9\%, highlighting the effectiveness of our generalization loss in addressing semantic bias.

\begin{table}[t]
\centering
\renewcommand{\arraystretch}{0.85}
\setlength{\tabcolsep}{1pt} 
\caption{
In-domain recommendation performance across four domains: IS, VG, MI, and Steam. Baseline models include GRU4Rec, SASRec, BERT4Rec, UniSRec, and RecFormer. Our \texttt{-Sem} variants consistently outperform base models, and \texttt{-RecG} achieves further significant gains by mitigating domain overfitting. Notably, \texttt{-RecG} outperforms all baselines across domains, demonstrating strong generalization and architecture-agnostic improvements.\vspace{-2pt}
}

\label{tab:indomain_performance}
\scalebox{0.95}{
\begin{threeparttable}
\small 
\begin{tabular}{lllllllll}
\toprule
\textbf{Model} & 
\multicolumn{2}{c}{\textbf{IS}} & 
\multicolumn{2}{c}{\textbf{VG}} & 
\multicolumn{2}{c}{\textbf{MI}} &
\multicolumn{2}{c}{\textbf{Steam}} \\
& R@10 & N@10 & R@10 & N@10 & R@10 & N@10 & R@10 & N@10 \\
\midrule
GRU4Rec         & 36.05 & 22.02 & 54.32 & 34.52 & 40.75 & 25.23 & 54.61 & 32.70 \\
SASRec          & 37.48 & 22.37 & 56.25 & 35.27 & 41.39 & 25.82 & 55.62 & 34.59 \\
BERT4Rec        & 39.09 & 23.62 & 53.50 & 32.97 & 45.80 & 28.74 & 56.49 & 35.68 \\
UniSRec         & 40.29 & 24.27 & 59.16 & 38.96 & 46.39 & 29.85 & 56.37 & 34.82 \\
RecFormer       & \underline{42.36} & \underline{25.16} & \underline{61.49} & \underline{40.28} & \underline{51.24} & \underline{31.65} & \underline{58.17} & \underline{36.92} \\
\midrule
GRU4Rec-Sem     & 44.45 & 26.17 & 62.11 & 40.59 & 51.66 & 31.58 & 58.35 & 37.55 \\
GRU4Rec-RecG    & 47.36 & 28.39 & 64.84 & 42.83 & 53.92 & \textbf{33.91*} & 59.62 & 38.14 \\
\midrule
SASRec-Sem      & 45.06 & 26.82 & 64.20 & 42.30 & 51.62 & 31.35 & 59.71 & \textbf{38.44*} \\
SASRec-RecG     & \textbf{48.57*} & \textbf{29.11*} & \textbf{65.92*} & \textbf{43.74*} & 53.47 & 33.73 & 60.39 & 39.87 \\
\midrule
BERT4Rec-Sem    & 44.81 & 26.47 & 63.55 & 41.34 & 51.25 & 31.14 & 59.74 & 37.28 \\
BERT4Rec-RecG   & 47.89 & 28.46 & 64.68 & 42.83 & \textbf{53.79*} & 32.96 & \textbf{61.63*} & 38.32 \\
\bottomrule
\end{tabular}
\end{threeparttable}
}
\vspace{-5pt}
\end{table}

\subsubsection{In-domain Sequential Recommendation Performance Comparison.}
In the in-domain scenario, we evaluate all baseline models, along with three base models each extended with two variants across four domains. The results in Table~\ref{tab:indomain_performance} yield several key observations: a) Variants augmented with LLM-based semantic information (\texttt{-Sem}) consistently outperform their corresponding base models, confirming the benefit of incorporating pretrained knowledge in domain-aligned tasks. For instance, BERT4Rec-Sem improves over BERT4Rec by up to 7.45\% in R@10 (VG domain).
b) Our proposed \texttt{-RecG} variants further outperform the \texttt{-Sem} models across all domains and models, with statistically significant gains. This highlights the effectiveness of our generalization loss in mitigating domain-specific overfitting due to the semantic gap between pretrained LLMs and the base recommendation models.
c) The performance gains of \texttt{-RecG} over \texttt{-Sem} are especially notable for SASRec, a unidirectional architecture that benefits more from the added generalization signal. For example, SASRec-RecG improves over SASRec-Sem by 3.51\% (R@10) and 2.91\% (N@10) in the IS domain.
d) Compared to state-of-the-art baselines (UniSRec and RecFormer), our \texttt{-RecG} variants achieve superior performance across all four domains. For instance, BERT4Rec-RecG achieves the highest R@10 in the MI and Steam domains with 53.79\% and 61.63\%, respectively, outperforming RecFormer by 2.55\% and 3.46\%.
Overall, these results demonstrate that our method not only improves in-domain effectiveness over traditional and text-only baselines but also provides a more general and robust framework applicable across different architectures.

\begin{table}[h]
\centering
\renewcommand{\arraystretch}{0.85} 
\setlength{\tabcolsep}{8pt} 
\caption{Ablation study on ZCDSR using BERT4Rec trained on IS and evaluated on VG and MI. Each component—item-level generalization (IG), intra-domain diversity (ID), inter-domain compactness (IC), and sequential-level generalization (SG)—contributes to performance. Removing ID causes the most significant degradation, highlighting its importance in preserving fine-grained domain distinctions. Overall, all modules collectively improve cross-domain generalization.\vspace{-2pt}
}
\label{tab:ablation_study}
\scalebox{0.95}{
\begin{threeparttable}
\begin{tabular}{lllll}
\toprule
\multirow{2}{*}{\textbf{Models}} & \multicolumn{2}{c}{\textbf{VG}} & \multicolumn{2}{c}{\textbf{MI}} \\
\cmidrule(lr){2-3} \cmidrule(lr){4-5}
                                 & R@10           & N@10           & R@10           & N@10           \\
\midrule
BERT4Rec-RecG                    & \textbf{35.81} & \textbf{21.53} & \textbf{30.10} & \textbf{17.06} \\
w/o IG                           & 32.49          & 18.36           & 26.49           & 14.83           \\
w/o ID                           & 28.26          & 16.33           & 20.93           & 12.17           \\
w/o IC                           & 32.74          & 19.79           & 27.35           & 15.77           \\
w/o SG                           & 33.17          & 19.22           & 28.63           & 16.52           \\
BERT4Rec-Sem                     & 31.84          & 17.78           & 24.82           & 13.28           \\
\bottomrule
\end{tabular}
\end{threeparttable}
}
\vspace{-5pt}
\end{table}

\subsection{Ablation Study(RQ2)}
We conduct an ablation study to analyze the impact of each generalization module under the ZCDSR setting.  Table~\ref{tab:ablation_study} shows the performance of BERT4Rec on VG and MI with IS as the source domain. The results highlight the role of item-level and sequential-level generalization in improving ZCDSR performance. Below, we summarize the findings for each variant:
 
 \textit{(a) Remove \underline{I}tem level  \underline{G}eneralization (IG).} When the item-level generalization is removed, the performance on both VG and MI drops noticeably (e.g., R@10 decreases from 35.81\% to 32.49\% on VG and from 30.10\% to 26.49\% on MI). This demonstrates the importance of IG in aligning item embeddings across domains and improving transferability.
 
\textit{(b) Remove  \underline{I}ntra-domain  \underline{D}iversity (ID).} Ablating intra-domain diversity (ID) causes the most significant performance degradation across all metrics, with R@10 dropping to 28.26\% on VG and 20.93\% on MI, even worse than \texttt{-Sem} variant. This emphasizes that maintaining fine-grained distinctions within each domain is crucial for preventing over-simplified representations, which can lead to poor generalization and reduced recommendation accuracy.

\textit{(c) Remove \underline{I}nter-domain  \underline{C}ompactness (IC).} Removing inter-domain compactness (IC) also results in a notable performance drop, particularly on VG (e.g., R@10 decreases from 35.81\% to 32.74\%). This indicates that aligning item embeddings across domains is essential for enabling effective knowledge transfer. However, the impact of removing IC is less severe than removing ID, suggesting that intra-domain diversity plays a more critical role in retaining domain-specific nuances.

\textit{(d) Remove  \underline{S}equential level  \underline{G}eneralization (SG).} Ablating sequential-level generalization (SG) reduces performance across both VG and MI, with R@10 dropping to 33.17\% and 28.63\%, respectively. This demonstrates that leveraging transferable sequential patterns is vital for capturing user behavior dynamics in the target domain. The smaller performance drop compared to IG and ID suggests that SG complements item-level generalization rather than acting as a standalone solution.

\begin{figure}[h]
    \centering  
    \subfigure{
        \includegraphics[width=0.231\textwidth]{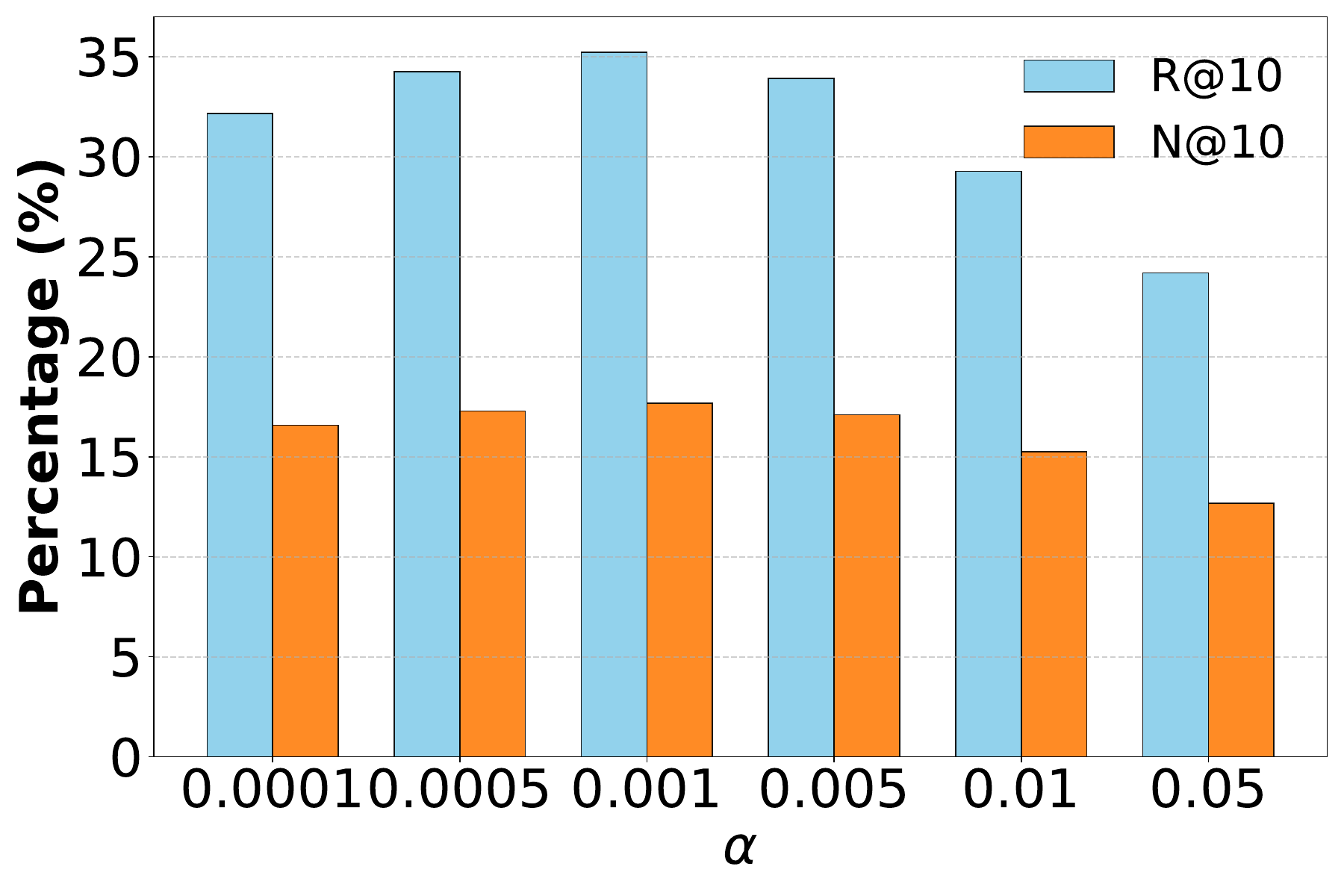}}
    \subfigure{
        \includegraphics[width=0.231\textwidth]{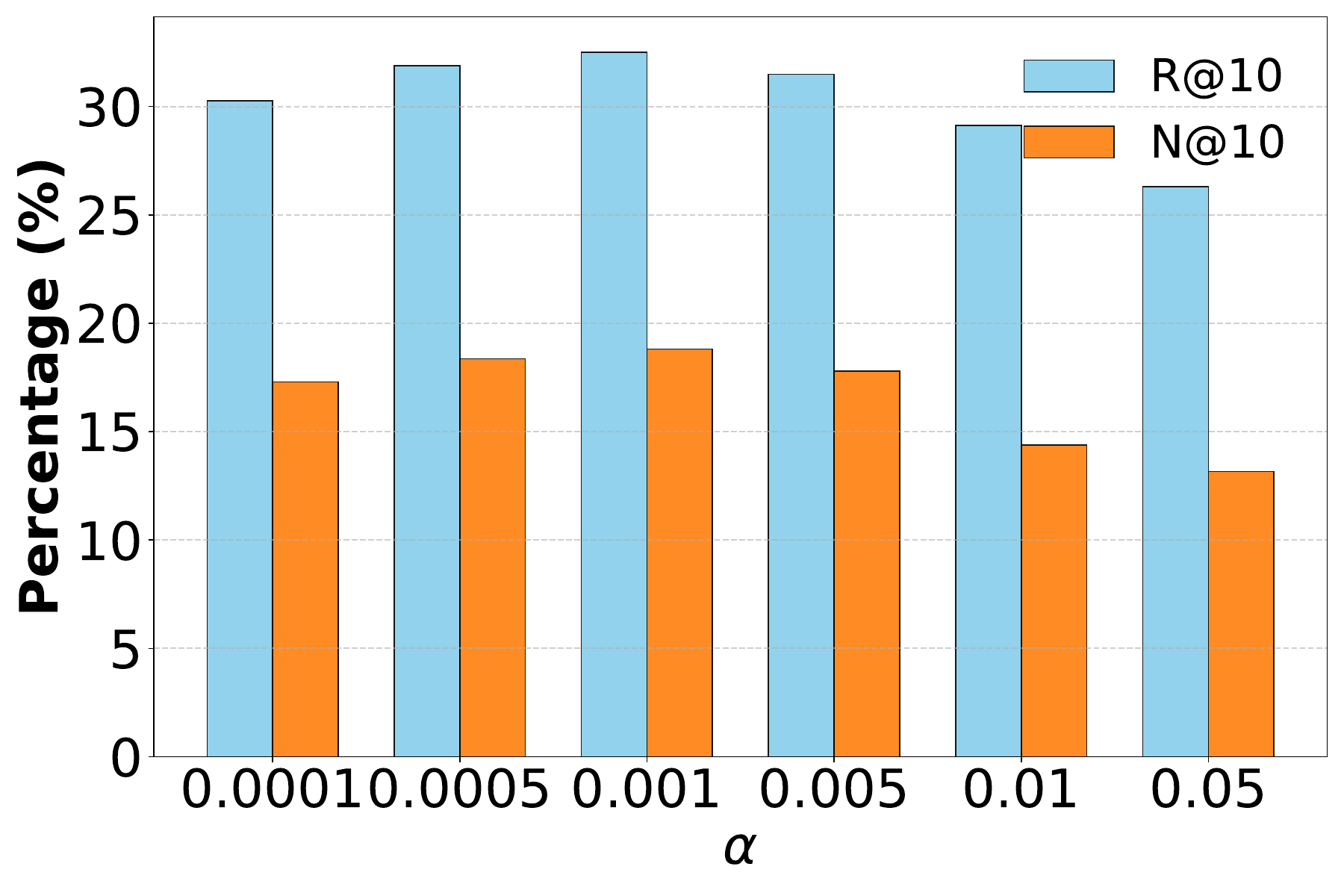}}
    \caption{
        Impact of the generalization weight $\alpha$ on ZCDSR performance over IS (left) and MI (right) datasets. A moderate value of $\alpha=0.001$ consistently yields the best performance, effectively balancing recommendation and generalization loss. Larger $\alpha$ values lead to performance degradation, highlighting the importance of tuning this parameter.
    }
    \label{fig:sensitivity}
\end{figure}
\vspace{-5pt} 

\subsection{Sensitivity Analysis(RQ3)}
To evaluate the impacts of key parameter $\alpha$ in ZCDSR tasks—we analyze its effect on the zero-shot performance of BERT4Rec-RecG across the IS and MI datasets. The hyperparameter $\alpha$ balances the recommendation loss and the generalization loss, making it a critical factor for optimizing ZCDSR performance. The results are visualized in \Cref{fig:sensitivity}, and the following observations are made:

A moderate value of $\alpha = 0.001$ achieves the best performance across both IS and MI domains, with significant improvements in Recall@10 and NDCG@10. This suggests that $\alpha = 0.001$ effectively balances the recommendation loss, which ensures accurate predictions, and the generalization loss, which aligns item embeddings across domains. However, as $\alpha$ increases beyond 0.001, performance declines consistently, indicating that overly large $\alpha$ values overemphasize the generalization loss, leading to suboptimal embeddings and reduced recommendation accuracy. The consistent trends across IS and MI further highlight $\alpha$'s robustness and its importance as a key parameter for optimizing generalization in ZCDSR tasks.

\begin{figure}[h]
    \centering

    \subfigure{
        \includegraphics[width=0.23\textwidth]{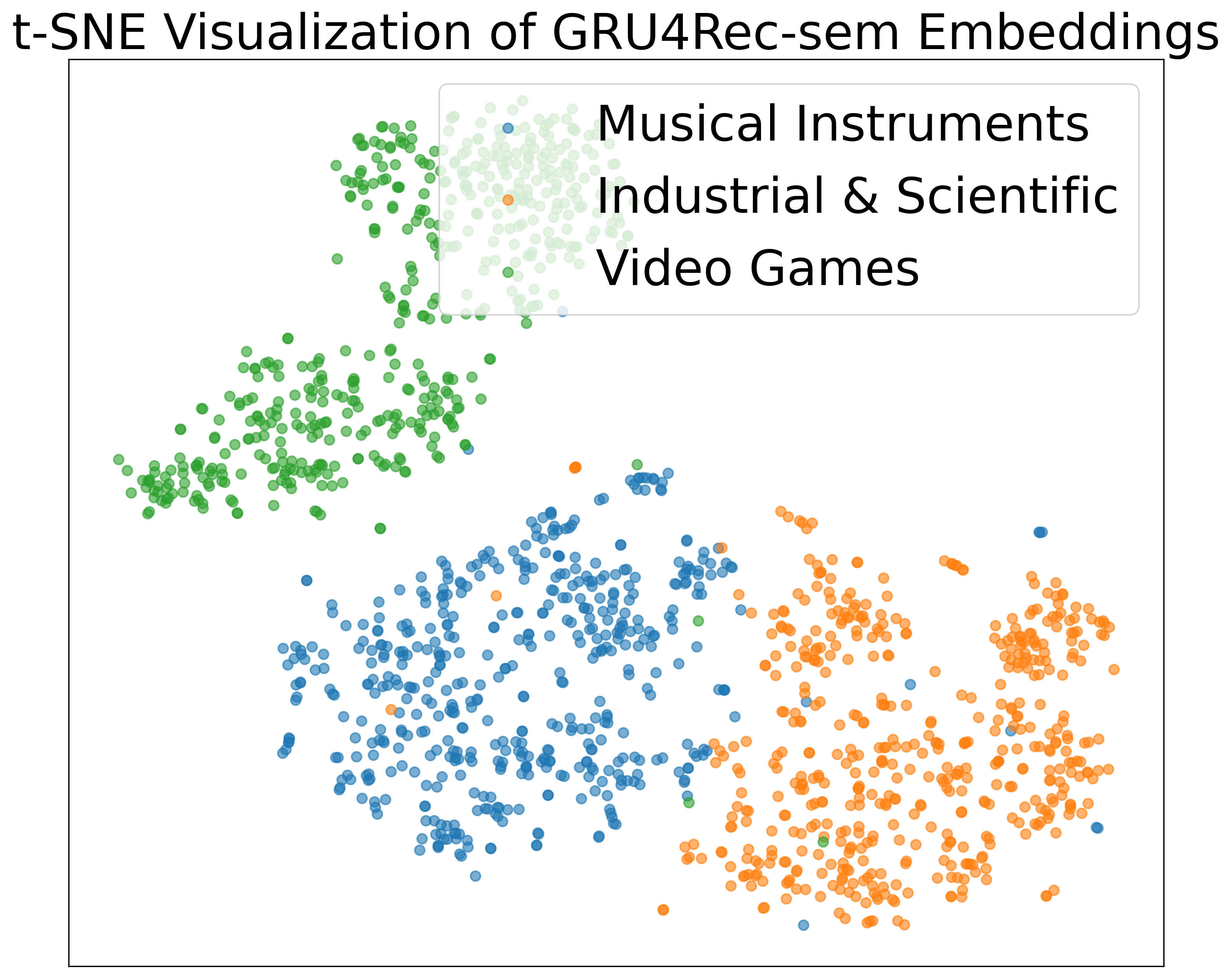}}
    \subfigure{
        \includegraphics[width=0.22\textwidth]{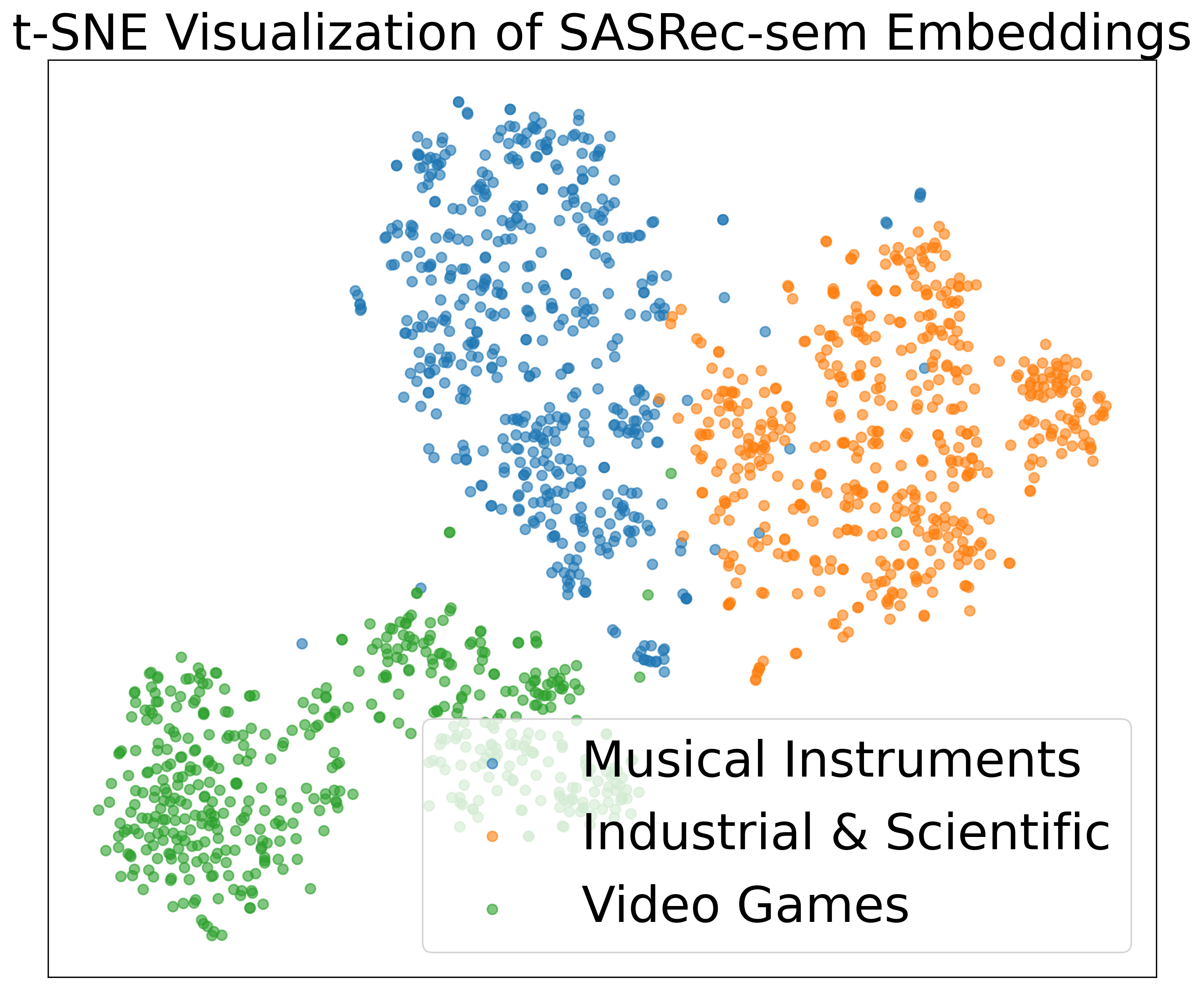}}
    \subfigure{
        \includegraphics[width=0.23\textwidth]{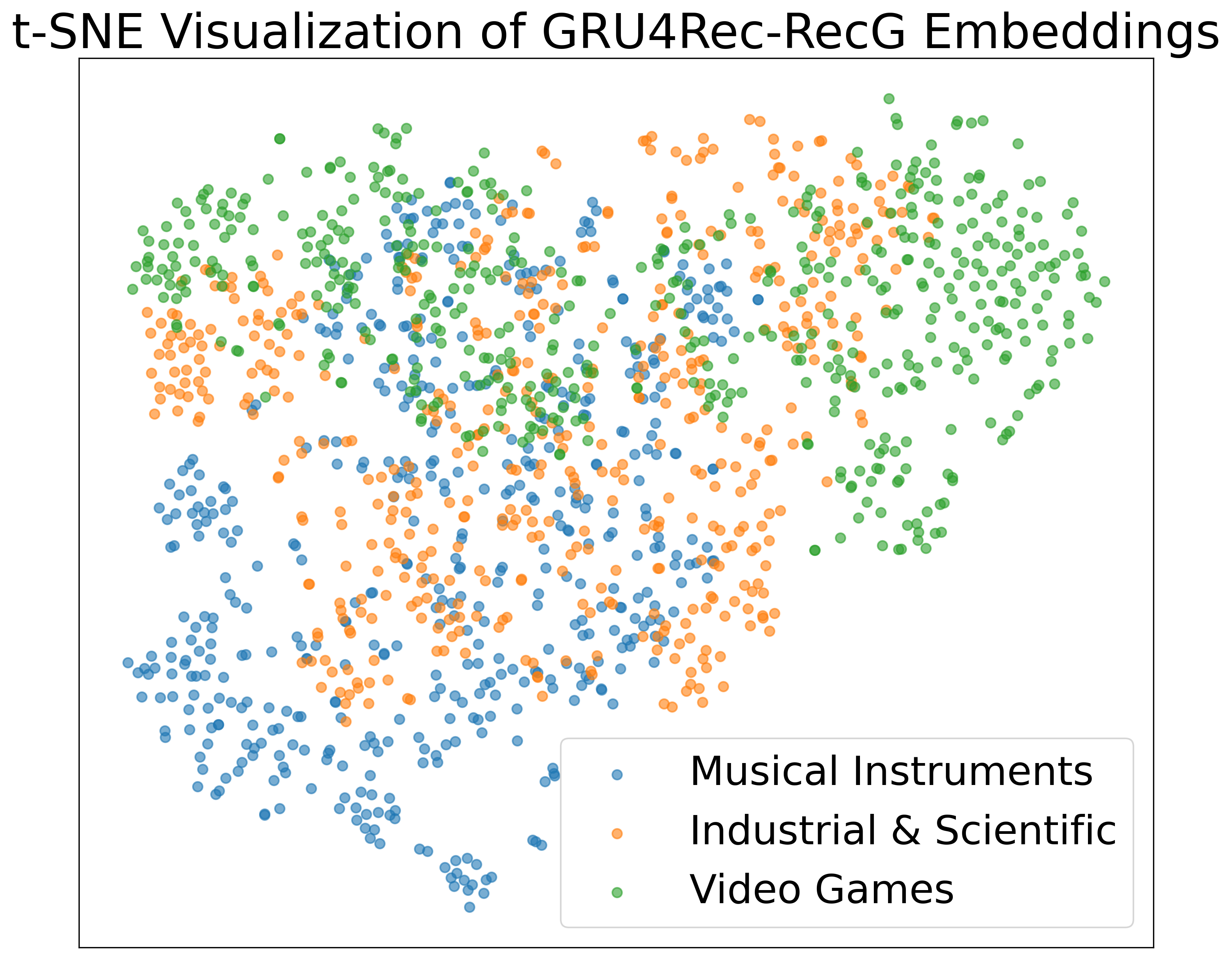}}
    \subfigure{
        \includegraphics[width=0.22\textwidth]{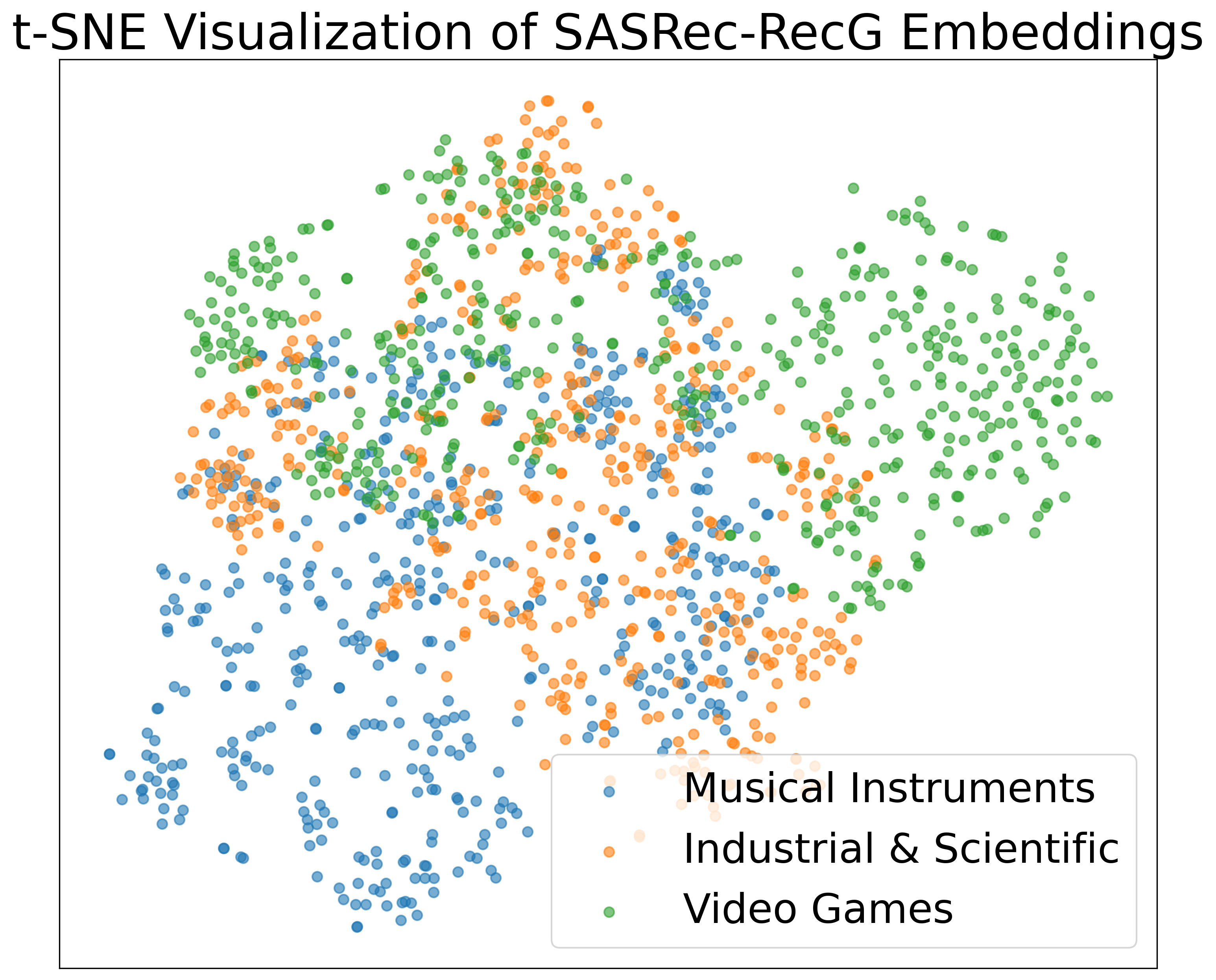}}

    \caption{
        Visualization of item embeddings on the Video Games domain: (a) GRU4Rec-Sem, (b) SASRec-Sem, (c) GRU4Rec-RecG, and (d) SASRec-RecG. While \texttt{-Sem} variants produce semantically distinct clusters, \texttt{-RecG} yields more uniformly distributed embeddings, improving generalization and robustness across domains.
    }
    \label{fig:vis}
\end{figure}

\subsection{Visualization Analysis(RQ4)}
To evaluate LLM-RecG affects the alignment and uniformity of item embeddings across domains—we analyze the item embeddings using t-SNE visualizations, as shown in ~\Cref{fig:vis}. These visualizations provide insights into the impact of our generalization framework on embedding alignment and generalization. The LLM-based semantic embeddings demonstrate a strong capability to distinguish items by leveraging prior knowledge. However, this strong semantic separation can hinder the generalization of the downstream recommender system. By incorporating the proposed generalization methods, the embeddings produced by the \texttt{-RecG} variants exhibit greater uniformity and are harder to distinguish across domains. This improved generalization enhances the model's robustness and adaptability.

\section{Related Work}

\subsection{Cross-Domain Sequential recommendation}
Unlike conventional sequential recommendation methods~\cite{kang2018self, sun2019bert4rec, li2021extracting, wang2021icpe} and cross-domain recommendation approaches~\cite{sheng2021one, luo2023mamdr}, cross-domain sequential recommendation (CDSR) requires models to leverage sequential dependencies to enhance recommendation performance across multiple domains. Early works, such as $\pi$-net~\cite{ma2019pi} and other approaches~\cite{Wang2020ACH, Yang2020LongSM, sun2021parallel}, employ recurrent neural networks to capture sequential information among overlapping users. More recently, $C^2$DSR~\cite{cao2022contrastive} introduces a contrastive Infomax loss with graph neural networks to capture both inter-sequence and intra-sequence item relationships. Unlike these methods, which assume full user overlap across domains, AMID~\cite{xu2024rethinking} devises a multi-interest debiasing framework capable of handling both overlapping and non-overlapping users. In this paper, we address a more general challenge: how to effectively utilize user interactions and item metadata to improve recommendation performance in scenarios where there is no overlap between users or items across domains. 

\subsection{LLM-based Recommendation}
LLMs' ability to generalize and understand text and long-term context has attracted interest in improving recommender systems. Research in this area is typically divided into generative and discriminative approaches based on LLMs' roles~\cite{wu2024survey}.

Generative recommender systems frame recommendation as natural language generation problems, enhancing interactivity and explainability. For instance, Rec Interpreter~\cite{yang2023large}, ChatRec~\cite{gao2023chat}, and He et al.~\cite{he2023large} use prompting strategies to convert user profiles or historical interactions into inputs for LLMs in conversational recommendations. ONCE~\cite{liu2024once} improves content-based recommendations by fine-tuning open-source LLMs and leveraging prompts with closed-source models. TALLRec~\cite{bao2023tallrec} fine-tunes Alpaca~\cite{taori2023stanford} using self-instruct data to provide binary feedback ("yes" or "no"). GenRec~\cite{ji2024genrec} takes a different approach by directly generating the target item for recommendation. Additionally, CCF-LLM~\cite{liu2024collaborative} encodes semantic and collaborative user-item interactions into hybrid prompts, fusing them via cross-modal attention for prediction.

Unlike generative systems, discriminative recommender systems do not require LLMs to respond during the inference stage, making them more practical and efficient. Early approaches, such as UniSRec~\cite{hou2022towards}, fine-tuned BERT~\cite{devlin2018bert} to associate item descriptions with transferable representations across recommendation scenarios. Harte et al.\cite{harte2023leveraging} improved sequential recommendation by replacing conventional model embeddings with LLM embeddings. LLM-ESR\cite{liullm} further enhanced sequential recommendation by integrating LLM-generated semantic embeddings via retrieval-augmented self-distillation, avoiding additional inference costs. Additionally, \cite{wang2024llm4msr,fu2023unified} leveraged LLMs to augment multi-domain recommendation models with scenario-specific knowledge, though their performance depends heavily on sample quality and domain expertise. In contrast, our method eliminates the reliance on domain knowledge, offering a more general and universal framework for cross-domain sequential recommendation.

\vspace{-5pt}
\section{Conclusion}
This work addresses the challenge of domain semantic bias in LLM-based zero-shot cross-domain sequential recommendation (ZCDSR), a critical task for achieving accurate predictions in unseen domains. We propose LLM-RecG, a model-agnostic generalization framework that mitigates domain semantic bias by operating at both the item and sequential levels. At the item level, it balances inter-domain compactness with intra-domain diversity to align item embeddings while preserving domain-specific nuances. At the sequential level, it transfers user behavioral patterns through clustering and attention-based aggregation, enabling dynamic adaptation without requiring target-domain interaction data.

Comprehensive experiments show that LLM-RecG consistently improves ZCDSR performance across diverse domains compared to the state-of-the-art CDR baselines, validating its effectiveness and scalability. Future directions include integrating richer metadata and addressing cold-start users to further enhance personalized zero-shot recommendation.

\newpage

\bibliographystyle{ACM-Reference-Format}
\bibliography{sample-base}








\end{document}